# Design and Implementation of Multilevel Access Control in Medical Image Transmission Using Symmetric Polynomial Based Audio Steganography

J.Nafeesa Begum
*Research Scholar &Sr. Lecturer in CSE, Government College of Engg, Bargur- 635104,*
*Tamil Nadu , India*
nafeesa_jeddy@yahoo.com

K. Kumar
*Research Scholar &Lecturer in CSE ,Government College of Engg, Bargur- 635104,*
*Tamil Nadu , India*
pkk_kumar@yahoo.com

Dr.V. Sumathy
*Asst .Professor in ECE , Government College of Technology,Coimbatore , Tamil Nadu,India*
sumi_gct2001@yahoo.co.in

*Abstract*— Steganography techniques are used in Multimedia data transfer to prevent adversaries from eaves dropping. The medical profession is also under a strict duty to protect the confidentiality of patients' medical records as per the HIPAA ACT of 1996. Thus, protection of medical data/information in telemedicine is of paramount importance. Most telemedicine systems include some form of security measures such as the use of passwords. Password identification determines whether a user is authorized to gain access to a system. However, passwords are insufficient mechanisms to maintain patient confidentiality from intruders who gain knowledge of a user's password to log onto a system. This paper deals with the design and implementation of multilevel access control in medical image transmission using symmetric polynomial based audio steganography. Medical image transmission using audio steganography deals with covering the medical image on top of an audio file and subsequently sending it to multiple receivers .Only the intended recipients know that the Audio files actually contain medical images and they can use their key to see the images. We have developed a Multilevel access control model by which medical images sent to low class users like medical assistants can also be seen by physicians who are higher in the hierarchy whereas the vice-versa is not allowed. To provide multilevel access control, symmetric polynomial based scheme is used. The steganography scheme makes it possible to hide the medical image in different bit locations of host media without inviting suspicion. The Secret file is embedded in a cover media with a key. At the receiving end the key can be derived by all the classes which are higher in the hierarchy using symmetric polynomial and the medical image file can be retrieved. The system is implemented and found to be secure, fast and scalable. Simulation results show that the system is dynamic in nature and allows any type of hierarchy. The proposed approach performs better even during frequent member joins and leaves. The computation cost is reduced as the same algorithm is used for key computation and descendant key derivation. Steganographic technique used in this paper does not use the conventional LSB's and uses two bit positions and the hidden data occurs only from a frame which is dictated by the key that is used. Hence the quality of stego data is improved.

*Index Terms*- HIPAA, Steganography, Multilevel Access control, audio file, symmetric polynomial, dynamic, scalable

## INTRODUCTION

The Health Insurance Portability and Accountability act (HIPAA) is widely acknowledged as the norm for healthcare services and Indian companies are well versed with the Act and other regulatory bodies. HIPAA [25] covers all protected healthcare information. It does not apply to specific records, but to information. The information is protected in any form by HIPAA and it continues to apply whether the content is being printed, discussed orally, or changes in form. For organizations that deal with the electronic management of healthcare information it is vital to protect the electronic maintenance and transmission of this data. Steganography is the art and science of writing hidden messages in such a way that no one, apart from the sender and intended recipient, suspects the existence of the message, a form of security through obscurity. The advantage of steganography, over cryptography alone, is that messages do not attract attention to themselves. Plainly visible encrypted messages will arouse suspicion, and may in themselves be incriminating in countries where encryption is illegal. Therefore, whereas cryptography protects the contents of a message, steganography can be said to protect both messages and communicating parties.

Steganography includes the concealment of information within computer files. In digital steganography, electronic communications may include steganographic coding inside of a transport layer, such as a document file, image file, program or protocol. Media files are ideal for steganographic transmission because of their large size.

There are many scenarios in which situation arises that only some users should be able to view the medical image or all higher level users should also be able to view the image message that is relayed to the doctors. for example in a orthopedic hospital all messages sent to surgeons should be seen by the pathologists, radiologists , anesthologists who are actually invisible doctors but play an important role in the operation by guiding the surgeon. There may be many other images sent to pathologists, radiologists, anesthologists that need not be known to the surgeons. Like wise all images sent to medical students need to be sent to the chief physician. There should also be a mechanism by which people at the same level are able to converse among themselves.

To implement such a multilevel access control in steganography symmetric polynomial approach is used. In most existing schemes, key derivation is different from key computation. Key derivation needs iterative computation of keys for nodes along the path from a node to its descendant, which is inefficient if the path is long. In this scheme, both operations are same by substituting (different) parameters in the same polynomial function assigned to node v. Thus, the key derivation efficiency can be improved. Our scheme also supports full dynamics at both node and user levels and permits any random access hierarchies. More importantly, removing nodes and/or users is an operation as simple as adding nodes and/or users in the hierarchy. A trusted Central Authority (CA) can assign secrets (i.e. polynomials) to corresponding nodes so that nodes can compute their keys. Also, nodes can derive their descendant's keys without involvement of the CA once polynomial functions were distributed to them. In addition, the storage requirement and computation complexity at the CA are almost same as that at individual users, thus, the CA would not be a performance bottleneck and can deal with dynamic operations efficiently.

The rest of the paper is as follows, Section 2 deals with related work Section 3 gives all overview of the system Section 4 describes the audio steganographic method for the images Section 5 deals about the symmetric polynomial approach Section 6 shows the simulation results and Section 7 gives the performance analysis and section 8 concludes the paper.

## II. RELATED WORK

Information hiding using steganography [9] relates to protection of text, image, audio and digital content on a cover medium [1,2,3,5].The cover media in many cases has been an image [1]. Aoki presented a method in which information that is useful for widening the base band is hidden into the speech data [6] .Sub band Phase shifting was also proposed for acoustic data hiding [3].All these schemes focus on data that is stored in a hard disk or any other hardware whereas there are many applications like military warfare where the audio data is to be given in real time as in live broadcast system. Techniques for hiding the audio in real time came into existence [4] and systems for synchronized audio steganography has been developed and evaluated [7]. In this scheme secret speech data is recorded and at the same time it is sent to the receiver and a trusted receiver extracts the speech from the stego data using the key which is shared between the server and the receiver. In the proposed scheme multilevel access control is implemented using symmetric polynomial approach. The audio is encrypted by a key of lower level user, and the higher level users are able to derive the key using symmetric polynomial approach. Forward and Backward secrecy [8] i.e. whoever is active at that instance only are able to receive the audio.

The first multi level access solution was proposed by Akl et al. [11, 12] in 1983 and followed by many others [13, 16, 17, 18, 20, 21, 14, 15, 19, 22, 23]. These schemes basically rely on a one-way function so that a node v can easily compute v's descendant's keys whereas v's key is computationally difficult to be computed by v's descendant nodes. Moreover, many existing schemes have some of the following problems: (1) Some schemes were found with security flaws in due course of time shortly after they were proposed; (2) Some schemes cannot support for reconfiguration of a hierarchy; (3) Some schemes require access hierarchy to be in a certain form so that it must be a tree or a DAG with only one root; and (4) Member revocation is one of the most difficult processes in cryptographic schemes, therefore, it is important to address this problem so that the revocation process can be dealt with efficiently.  In this paper, we propose a new scheme based on symmetric polynomials for synchronized audio data. Unlike many existing schemes based on one-way functions, our scheme is based on a secret sharing method which makes the scheme unconditionally secure [21, 24]. Also, this multilevel access control requires two types of key operations: (1) key computation: a node v computes its own key, and (2) key derivation: a node v computes its descendants' keys.

## III. SYSTEM OVERVIEW

Multilevel Access Control in medical image transmission is useful for Hospitals which have a hierarchical structure. For e.g. In a Orthopedic hospital, many people are involved at various levels and there is a hierarchy among them as shown below.

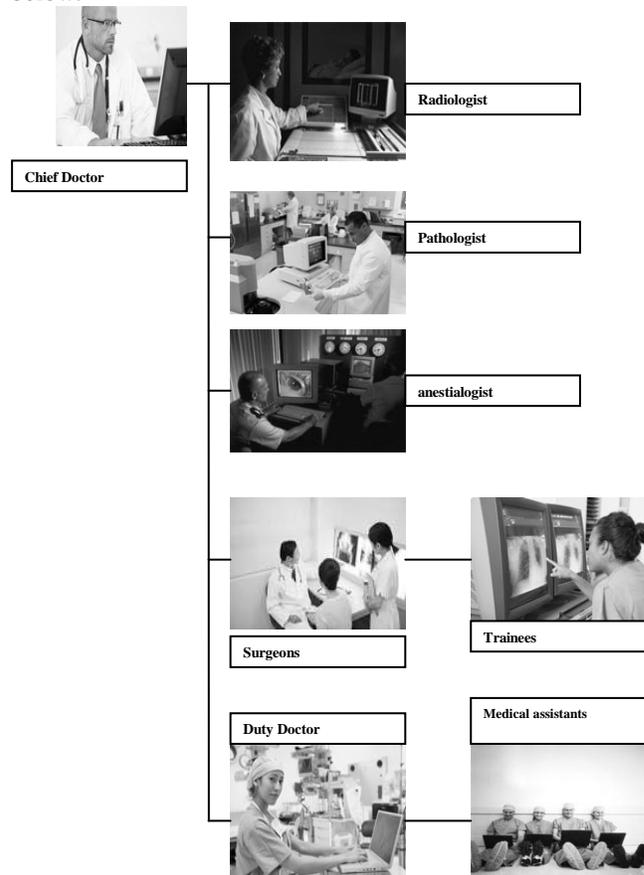

Figure: 1 : A sample hierarchy in a medical organization

In such a type of system, medical images sent to a lower class should be seen by the active members of lower class and also by all active members of the higher class. It is not only essential to maintain the access control but, the data should be hidden as well. Steganography technique is used to hide the image into audio file and sent .The sequence of events is as follows.

*At the server*:

1. Generate a general polynomial.

2. Give a symmetric polynomial to each of the classes.

3. Select the image to be sent

4. Use Steganographic technique to hide the image into audio.

5. A text can also be hidden in a audio file.

6. The file is encrypted by the class key for whom the message is to be relayed.

7. The symmetric polynomial generates a key in this case.

8. The server takes care of including class dynamics so the hierarchy can be changed at any time.

9. Users can join or leave a class at all instances. Keys are recalculated so that Forward and Backward secrecy are maintained.

10. If the users within the group need to transfer message among themselves, the private key of the users is used.

At the receiver

1. All the active receivers will receive the audio file.

2. If the recipient belongs to the actual intended class he can use the polynomial to get the hidden medical image file instantaneously from the audio file.

3. If the recipient belongs to a class lower than the actual intended class in the hierarchy, he will not be able to derive the key .The polynomial derivation method will give a null value.

4. If the recipient belongs to a higher class he can derive the key of the lower class and see the image file which in turn is applicable to text messages as well.

5. The users at the same class can transfer messages among them.

6. When a user joins or leaves, the new polynomials are given by the server and the private keys also get updated according to the new polynomial. Other classes are not affected by this.

7. Service messages can be sent from higher class users to lower class users.

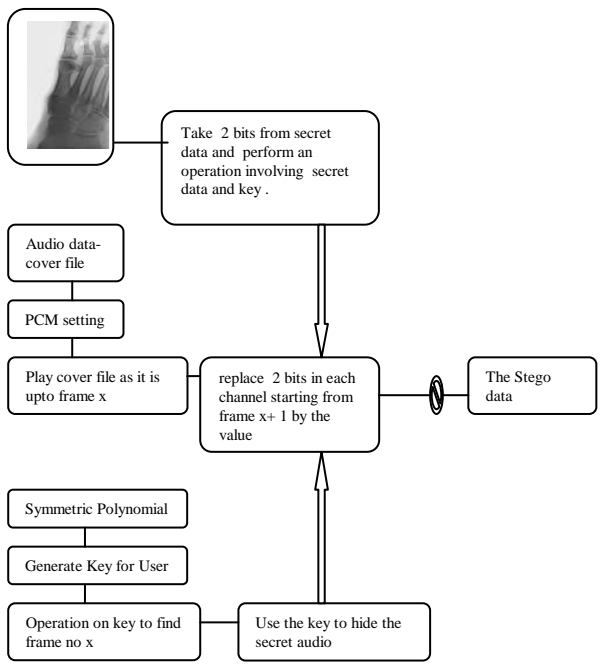

Figure 2: At the Server

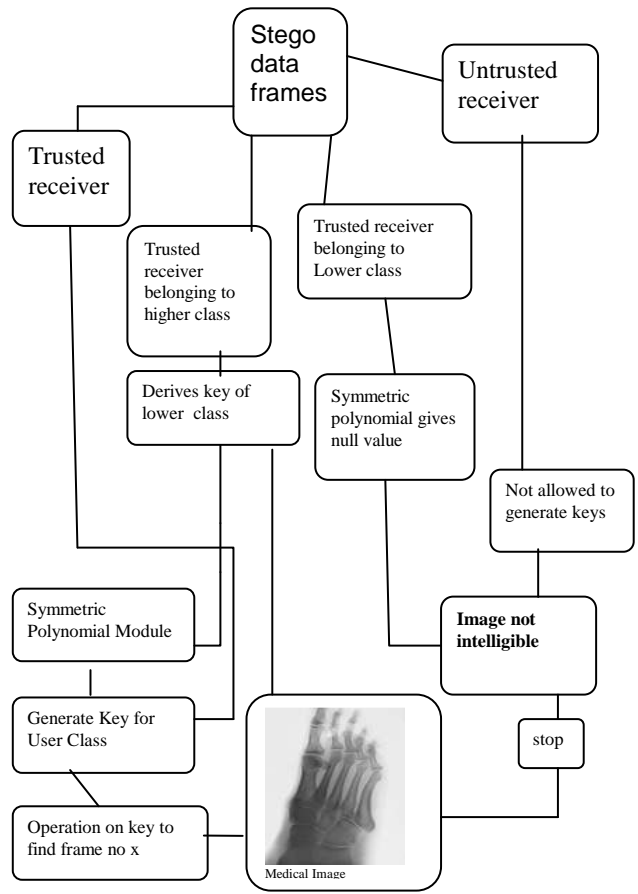

Figure 3: At the Receiver

### 3.1. Contribution

Solutions of a synchronized steganography have been given in past [7]. In this once the stego data reaches the destination the audio can be listened by the trusted receiver. Our contributions are

1. The key is used during the embedding process also.
2. The key is not a simple key but it identifies a class of users.
3. If the key used belongs to a low level group in the hierarchy, The higher level class of user can derive the key using the symmetric polynomial approach and listen to it.
4. There can be normal message transfer among the group elements and also service messages from higher classes.
5. Forward and backward secrecy is maintained.
6. It is a dynamic one where new hierarchies can be introduced, User level and class level dynamics are taken care.

### IV AUDIO STEGANOGRAPHY FOR MEDICAL IMAGES

The input Image to be sent is selected. When the covering media is being played at the same time the image is can be put into the audio file simultaneously or the image can as well be hidden in a already stored file. The stego bit stream is then transmitted to the receivers. Multilevel Access control using symmetric polynomial is used at this stage to generate the key to make secure transmission of the image file. According to the hierarchy the trusted users are able to retrieve the hidden audio file.

In this system, cover data is divided into fix-sized frames according to Pulse Code Modulation setting. To cover low size and high phonetic quality suitable sampling rate for the cover media is selected. Three main processes are involved in the Audio Steganography for Medical Images.

**(1).** Using data sampling    image data's are embedded into another audio.

**(2).** Bit Embedding: The key used helps in hiding the image file in bit positions and once the bit positions are found data is hidden after performing an operation on secret data and the key.

**(3).** Synchronized Process: Malicious and intentional attacks can be avoided as the secret data can be put over a real time audio input also.

### 4.1 Algorithm

Step1: select the image file. The audio file is divided into fix-sized frames and set to be specific PCM format. PCM qualification is decided by sampling rate, sampling size, and sampling channel. The PCM property of cover audio wav is set to be 32 kHz-16bit-2ch.

FORMULA USED FOR STEGANOGRAPHY:

Steganographic process:
cover medium + hidden data + stego_key = stego_medium
Part of The Wave File Format opened using Notepad:

```
52 49 46 46 24 40 01 00  57 41 56 45 66 6D 74 20  RIFF$@..WAVEfmt
10 00 00 00 01 00 02 00  11 2B 00 00 44 AC 00 00  .........+..D...
04 00 10 00 64 61 74 61  00 40 01 00 00 00 00 00  ....data.@......
```

The format is being specified by a WAVEFORMATEX structure

| 01 00 | 02 00 | 11 2B 00 00 |
|---|---|---|
| WAVE_FORMAT_PCM | count of channels | samples per second |

| 44 AC 00 00 | 04 00 | 10 00 |
|---|---|---|
| bytes per second | block align | bits per sample |

Step2:  Use Symmetric Polynomial to calculate key of class
Step3: Perform calculation and decide the frame from which the data is to be embedded.
Step 4: Decide two bit locations in each frame and clear the bit in the locations $cmask1 = (2^{loc1} - 1)$ xor (Keybit),
$cmask2 = (2^{loc2} - 1)$ xor (Keybit), $cmask = cmask1 \wedge cmask2$ hide the secret data bits into these bit locations by again performing an operation on the secret data along with the key. The cover media has two channels and so the data is written on both the channels. Other bits are not changed.
Step5: The next set of data will go to the next frame.
Step6: Do the repetitive process till the entire image is hidden.
Step8: Transmit using sockets
Step9: At the receiving end, Use the key and see the image
Step10: If the receiving user belongs to higher class, he can derive the key and listen to the audio.

### V SYMMETRIC POLYNOMIAL APPROACH PROTOCOL

A polynomial F(x; y; z) is said to be symmetric if F(x; y; z) = F(x; z; y) = F(y; x; z) = F(y; z; x) = F(z; x; y) = F(z; y; x):A polynomial in several variables $x_1, \dots x_n$ is called    symmetric if it does not change when you permute the variables.

5.1 A Symmetric Polynomial Based Multilevel Access Scheme

- A is an set of n classes – { C1,C2,C3,......,Cn}
- B is a set of ancestral classes of set A.

  B = {S1,S2,S2,.......,Sn}

- mi is calculated as the number of the ancestral classes

  mi = |Si|

- We are choosing a value for m , such that m ≥ max {m1,m2,m3,........ mn} + 1

- Here m is the number of parameter in the polynomial function P, where P is for constructing  our multi level access control scheme

A Numerical Example:

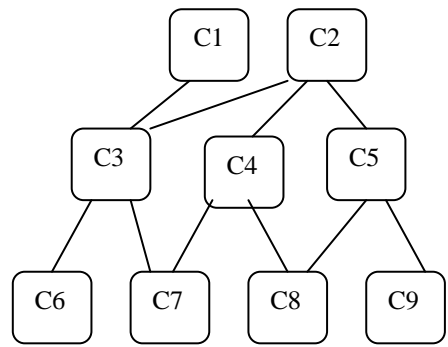

Figure 4: Hierarchical Structure of Multi Level access control

- Here we have nine classes
  { C1,C2,C3,C4,C5,C6,C7,C8,C9}
- Ancestral classes sets are
  S1 = { φ }
  S2 = { φ }
  S3 = { C1,C2 }
  S4 = { C2 }
  S5 = { C2 }
  S6 = { C1,C2,C3 }
  S7 = { C1,C2,C3,C4 }
  S8 = { C2,C3,C5 }
  S9 = { C2,C5 }
- From the previous step, we need to choose m such that
  m ≥ max { m1,m2,m3,….,m9} + 1
      Let us choose m=7, It will allow us to expand the hierarchy without changing the value of m.
- Symmetric polynomial, we are using here is as follows

$$P(x1,x2,....,xm) = \sum_{i1=0}^{t} \sum_{i2=0}^{t} \sum_{im=0}^{t} a_{i1,i2,..im} \, x_1^{i1} x_2^{i2}.x_m^{im} \, (mod \, p)$$

  Here t is threshold number.
- We can classify the work into two types.
  i) Key Calculation
  ii) Key Derivation

**Key Calculation**
- We can calculate key $K_i$ of class $C_i$ as follows

$Ki = P(S_i, S_{i1}, S_{i2}, ....., S_{im}, S'_1, S'_2, ....., S'_{m-mi-1})$     - (1)

**Key Derivation**
- In key derivation, we are using a term $S_j/I$ which can be calculated as

  $S_{j/1} = S_j/(S_i \cup \{Ci\})$
        $= \{ C_{(j/i)1}, C_{(j/i)2}, ........., C_{(j/i)rj} \}$
- Consider a class $Ci$ which is ancestor to class $Cj$ and key $Kj$ can be calculated by $Ci$ as ,

  $K_j = g_i(S_j, S_{(j/i)1}, S_{(j/i)2}, ............, S_{(j/i)rj}, S'_1, S'_2, ....., S'_{m-mi-2-rj})$

  $= P(S_i, S_j, S_{i1}, S_{i2}, ..........., S_{imi}, S_{(j/i)1}, S_{(j/i)2}, ............, S_{(j/i)rj}, S'_1, S'_2, ....., S'_{m-mi-2-rj})$ (2)

**Example**
**Key Derivation**
      Consider that C3 is an ancestor class to class C7. Then K7 can be derived by C3 in the following steps.

        $S_{7/3} = \{ C4 \}$
        $K7 = P (S3,S7,S1,S2,S4,S1',S2')$

General Concept used in the polynomial:
    for(i=0;i<t;i++)
    for(j=0;j<t;j++)
    for(k=0;k<t;k++)
    for(l=0;l<t;l++)
    for(m=0;m<t;m++)
    for(n=0;n<t;n++)
    for(o=0;o<t;o++)
    {        $a[o] = o^2 + 10$
            $a[o] = a[o]*a^i.b^j.c^k.d^l.e^m.f^n.g^o$
            $c = (c+a[o]) \% s$
    }

**Key Calculation for the Classes using equation (1)**
K1 = P(S1,S1',S2',S3',S4',S5',S6')
K2 = P(S2,S1',S2',S3',S4',S5',S6')
K3 = P(S3,S2,S3, S1',S2',S3',S4')
K4 = P(S4,S1,S2,S1',S2',S3',S4')
K5 = P(S5,S1,S2,S1',S2',S3',S4')
K6 = P(S6,S1,S2,S3,r1,r2,r3)
**K7 = P(S7,S1,S2,S3,S4,S1',S2')**      -(3)
K8 = P(S8, S1,S2,S3,S4,S5,r1)
K9 = P(S9, S1,S2,S3,S4,S5,r1)

**Key Derivation of class 7 by class 3 using equation(2)**
S3 = { C1,C2 }
S7 = { C1,C2,C3,C4 }
S3∪{ C3 } = { C1,C2,C3 }
S7/3 = { C4 }
**K7 = P( S3,S7,S1,S2,S4,S1',S2' ) – (4)**
      Which is equal to the key calculated by class7 itself.
**Key Derivation of class 3 by class 7 using equation(2)**
S3 = { C1,C2 }
S7 = { C1,C2,C3,C4 }
S7∪{ C7 } = { C1,C2,C3,C4,C7 }
S7/3 = { φ }
**K7 = P( S7,S3,S1,S2,S3,S4,S1' ) –(5)**

It can be seen that when the class derives its own key  and when an ancestor of this class derives the key , equation 2 and equation 4  have same parameters  passed in the polynomial but the combination differs whereas when a wrong ancestor derives the key the parameters are not the same.
The default values, we have taken are
1)  m=7

2)  P=2147483646

3)  s1=5, s2=10,s3=13,s4=9,s5=6,s6=22,s7=18,s8=30, s9=39,

r1=11,r2=12,r3=13,r4=14,r5=15,r6=16,r7=17,r8=18, r9=19 (instead of s' we have used r )

For a small Hierarchy, with more than two classes, we can easily illustrate our key calculations ,where each class consists of several users.

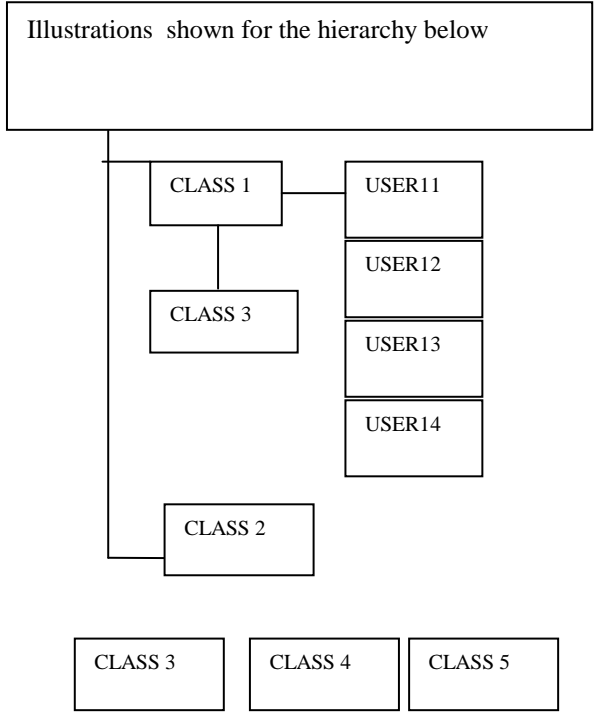

Figure 5:Normal Structure

**Key Calculations:**

**Step-1:**

The parameters to be passed for calculating group key class 2 are

P(s2,r1,r2,r3,r4,r5,r6)

Group key class C2

    =699615258

1) Private key for user21

        =699615258+(699615258/21)

        =732930270

2) Private key for user22

        =699615258+(699615258/22)

        =731415951

3) Private key for user23

    =699615258+(699615258/23)

        =730033312

4) Private key for user24

    =699615258+(699615258/24)

        =728765893

**Step-2:**

The same way group key for class 4 can also calculated by passing the following parameters to the polynomial

P(s4,s2,r1,r2,r3,r4,r5)

  Group key for class 4=1947982264

1)Private key for user41

        =1947982264+(1947982264/41)

        =1995494026

2) Private key for user42

        =1947982264+(1947982264/42)

        =1994362794

3) Private key for user43

        =1947982264+(194782264/43)

        =1993284177

4) Private key for user44

        =1947982264+(194782264/44)

        =1992254588

**Key Derivation:**

Deriving the group key of class C4 using its ancestral class C2

S2={}

S4={s2}

Sj|I can be calculated as

S4|2={s2}|(s2U{})

={s2}\{s2} Set Difference

S4|2=O

The parameters to be passed for deriving the key of class C4 using C2

        M-mi-2-rj

Key=p(s2,s4,r1,r2,r3,r4,r5)

    =p(10,9,11,12,13,14,15)

    =1947982264

Private Keys are used for local communication among the users.

VI.SIMULATION RESUTS

The system is developed using .NET and found to be secure and fast. The system takes care of User level and Class level dynamics. The large number of numbers prevents a possible guessing .for eg. For a eight parameter polynomial, 16! (i.e. 10922789888000) combinations possible. Bursty leave and join operations also are possible and the system can be used for any hierarchy.

The additional features that are included are

1.message passing among the same group.

2.service messages from higher users to lower level users.

3.Dynamics at user level and class level and secrecy.

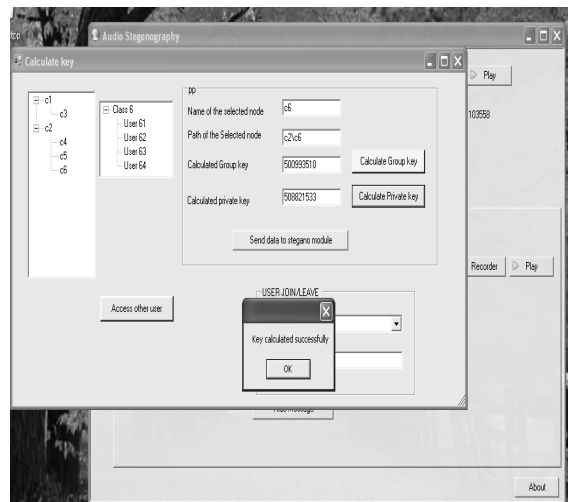

Figure 6: Key calculation

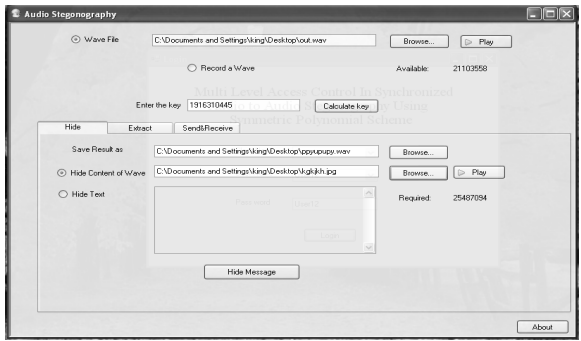

Figure 7: Hiding the image in Audio

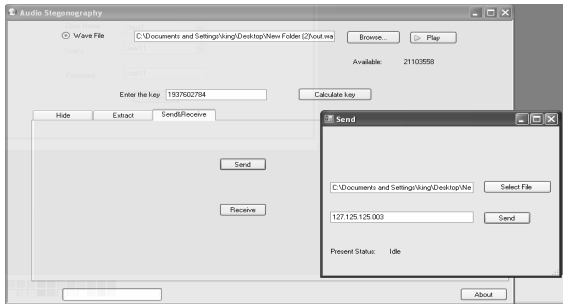

Figure 8: sending the image

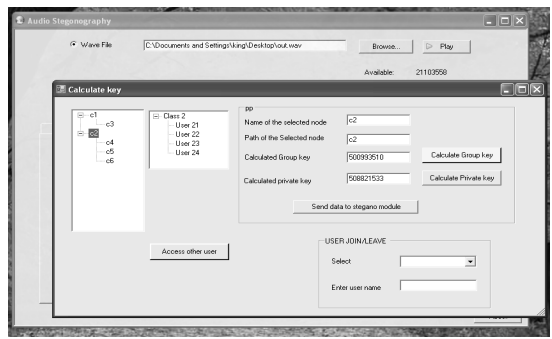

Figure 9: Deriving the Key

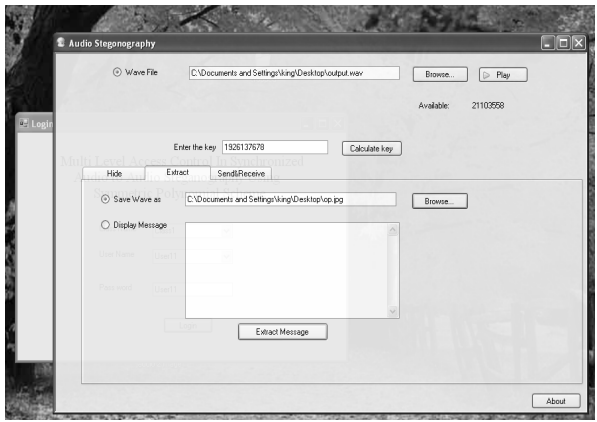

Figure 10.Extracting the image file

## VII PERFORMANCE ANALYSIS:

Performance and security : Each user $u_i$ will receive the time complexity for computing the group key is $O(w^{2n})$. An important performance measure for a secure group communication scheme is the number of rekeying messages. Suppose that t users will be joining the group. The TA will send k and $g_i$ to each of them respectively (2t messages) and broadcast one message to tell which users are joining. The total number of rekeying messages is O(2t). Suppose that t users are leaving the group. The TA only broadcasts one message to tell which users are leaving, thus the number of rekeying messages is O(1). Suppose that t users are joining and another v user is leaving the group, the total number of rekeying messages is still O (2t).

As for the security of the scheme, if w +1 users collude, then they can figure out the function f entirely. Therefore, the scheme is w-resilient. Moreover, if less than w + 1 users collude, they cannot get any information about the key, i.e., any value in the key space looks like a valid and equiprobable key to these colluding users. It follows that

$$g_i = f\left(s_1, x_2, \cdots, x_n\right) = g_i(x_2, \cdots, x_n) = \sum_{i_2=0}^{i_2=w} \cdots \sum_{i_n=0}^{i_n=w} a_{i_2, \cdots, i_n} x_2^{i_2} \cdots x_n^{i_n}.$$

the scheme is unconditionally secure.

*A. Memory:* Each user will be able to calculate the key based on the polynomial and hence very less memory is used. All parameters are publicly available and using the same method keys of lower hierarchy can be derived by substituting the corresponding parameters as given by the derivation module. The steganographic module does not involve any storage for storing the already recorded data as always data is recorded and subsequently sent to the receivers. The size of the cover medium does not increase because only two bits are used in each channel.

*B. Computation cost:* When the user joins there is no need for recalculation because the recorded message has already been placed. When a user leaves a group key is recalculated and given to the class. Private keys are generated from this. The value of the new key involves not a change of parameters but a change of mod P value. Hence each class will be able to get a new polynomial value by passing the same parameters. Any left user will not be able to get the key. Only one key is used during creation of stego data. The higher class users need not remember the keys of all their descendant classes but rather using a simple scheme derive the exact parameters to generate the key. Hence the computation cost is reduced.

*C. Communication Cost:* The p value is changed by the Trusted Authority and when the users try to calculate the key the new key will be generated. The computation cost is reduced because, the class users are not bothered about the key transmission. Once the polynomial is given the users can calculate their own key.

*D. Dynamics:* Class joining, Class leaving, Dynamic Hierarchy and New user joining a class are all done by the trusted authority in a phased manner thereby allowing the scheme to scale to greater hierarchy .Additionally local messaging and service messages are taken care in this system.

## VIII.CONCLUSION AND FUTURE WORK

Thus in this paper , Design And Implementation of Multilevel Access Control in Medical Image Transmission Using Symmetric Polynomial Based Audio Steganography    is done. The implementation results show that any type of hierarchy can be introduced and all dynamics can be done. For a 8 parameter polynomial that can have 8 among the 16 values, there are 16 ! combinations. Also, nodes can derive their descendants' keys without involvement of the CA once polynomial functions are distributed to them. In addition, the storage requirement and computation complexity at the CA are almost same as that at individual nodes, thus, the CA would not be a performance bottleneck and can deal with dynamic operations efficiently.

We believe that this flexible, integrated and easy to use solution is a robust alternative to more complex architectures for simple image transmissions or occasional circumstances. In many fields (e.g., medicine, manufacturing, veterinary science, scientific research, etc.), it is often necessary to examine a subject and communicate the results of the examination to a remote place. Such information exchanges are especially desirable in the medical arena where it is often useful for medical practitioners to communicate medical information, such as patient test results, e.g., radiology studies or cardiac studies, to other practitioners located in remote places. This type of system provides secure transmission with multilevel access control.

*8.1 Future work*

1.The Trusted authority can still make it secure by changing the value of the parameters .

2. Should be extended for video files.

3. Bit selection for steganography can be made by using some pseudo random generator.